# Suppression of the critical temperature of superconducting NdFeAs(OF) single crystals by Kondo-like defect sites induced by α-particle irradiation

C. Tarantini[1], M. Putti[3], A. Gurevich[1], Y. Shen[2], R.K. Singh[2], J.M. Rowell[2], N. Newman[2], D.C. Larbalestier[1], Peng Cheng[4], Ying Jia[4], Hai-Hu Wen[4]

[1]National High Magnetic Field Laboratory, Florida State University, Tallahassee, FL, USA

[2]Department of Materials Science and Engineering, Arizona State University, Tempe, AZ, USA

[3]CNR-INFM-LAMIA and Department of Physics, University of Genova, Genova, Italy

[4]Institute of Physics and National Laboratory of Condensed Matter Physics, Chinese Academy of Sciences, Beijing, China

**Abstract**

We report a comprehensive investigation of the suppression of the critical temperature $T_c$ of NdFeAs(OF) single crystal by α-particle irradiation. Our data indicate that irradiation defects produce both nonmagnetic and magnetic scattering, resulting in the Kondo-like excess resistance $\Delta\rho(T) \propto lnT$ over 2 decades in temperatures above $T_c$. Despite high densities of irradiation defects, the dose at which $T_c$ is suppressed to zero is comparable to that for $MgB_2$ but is well above the corresponding values for cuprates.

PACS: 74.25.Fy, 61.80.Jh, 74.62.Dh.

Superconductivity in ferropnictides has attracted much interest because of high critical temperatures $T_c$ and the unusual interplay of multiband superconductivity and antiferromagnetism (AF) mediated by the magnetic Fe ions in semi-metallic AF parent compounds, which have spin density wave transitions at 130-150K [1]. In rare-earth pnictides REOFeAs (RE =Ce, Pr, Nd, Sm), magnetism can also manifest itself in the rare-earth 4f states causing competition between superconductivity, paramagnetism, AF and the Kondo effect [2]. It has been suggested by many authors that superconductivity in pnictides may result from a superexchange repulsion mediated by magnetic excitations, which couple electron and hole pockets of the Fermi surface. Such pairing interaction favours either isotropic s-wave order parameters with opposite sign on different sheets of the Fermi surface ($s^{\pm}$ model) or anisotropic s-wave or d-wave order parameters with nodes [3,4]. Experimental confirmations of these scenarios remain inconclusive. Measurements of the London penetration depth $\lambda(T)$ have shown both exponential temperature dependencies consistent with a fully gapped Fermi surface [5] and power law behaviours indicative of nodal quasiparticles [6] or impurity-induced bound states at the Fermi level [7]. The response of the superconducting state to magnetic and nonmagnetic impurities is thus one of the key manifestations of different pairing symmetries in pnictides.

The effect of impurity scattering could in principle be revealed by measuring $T_c$ as the concentration of impurities is varied. Recently it was shown that Co, Ni and Zn substitutions on Fe sites only weakly suppress $T_c$ [8]. However, such experiments are often complicated by substitutional chemistry, inhomogeneous distribution of impurities or precipitation of second phases, which makes it difficult to incorporate high densities of impurities. Impurities can also result in doping effects, shifting the chemical potential and masking the effect of disorder of $T_c$. Under these circumstances, irradiation becomes the most straightforward way of introducing disorder without doping effects. However, because pnictides are built of magnetic ions, their displacements by irradiating particles can produce an unusual type of point defects which cause both nonmagnetic and spin-flip scattering. The magnetic component may come from disruption of nearest neighbour Fe and As orbitals and a partial restoration of the magnetic moment of the Fe ion [9]. This mechanism may be enhanced by displacing the magnetic RE ions. Thus, irradiation of pnictides enables a probe of the effect of both nonmagnetic and magnetic scattering on $T_c$ caused by the same defects. So far there has been only one report showing the suppression of $T_c$ by neutron irradiation of a polycrystalline LaFeAs(OF) followed by recovery of $T_c$ after annealing [10]. In this Letter we report resistivity measurements of NdFeAs(OF) single crystal in which point defects have been introduced in a controlled way by α-particle irradiation. As $T_c$ is progressively suppressed by irradiation, the resistivity curves develop a low temperature upturn which becomes more pronounced as the dose increases. This upturn suggests the Kondo-type spin flip scattering by irradiation defects, which however cause an unexpectedly weak $T_c$ suppression even for strong magnetic and nonmagnetic disorder.

Our experiment was performed on a NdFeAsO$_{0.7}$F$_{0.3}$ single crystal (Nd-1111) grown by a multistep flux process [11]. Pt contacts for resistivity and Hall measurements were made by the Focused Ion-Beam. The thickness of the sample was about 1 μm and $T_c$ before irradiation was 46.4 K. The α–particle irradiation at 300K was carried out using a 2 MeV $^4$He$^{2+}$ ion beam from a Tandem accelerator. A total dose of 5.25×10$^{16}$/cm$^2$ was achieved after 14 steps of irradiation. Simulations using the Stopping and Range of Ions in Matter-2003 software have shown that the mean free path of $^4$He$^{2+}$ ions in NdFeAsO$_{0.7}$F$_{0.3}$ is about 4.2 μm, which ensures uniform radiation damage throughout the sample, the collisions occurring mainly on the Nd, Fe and As sites.

Figure 1 shows that $T_c$ decreases monotonically after each irradiation step without significant broadening of the transition, and superconductivity was completely suppressed after the accumulated dose of 5.25×10$^{16}$/cm$^2$. Resistivity $\rho(T)$ progressively increased after each irradiation dose with a significant upturn developing at low $T$. Such evolution of $\rho(T)$ caused by irradiation is different from what has been observed on other superconductors: in MgB$_2$ the residual resistivity $\rho_0$ increases without any upturn at low temperatures [12,13] while in cuprates a low temperature upturn $\Delta\rho(T)$ develops only after much higher doses [14,15]. The observed $\Delta\rho(T) \propto lnT$ in cuprates caused by Zn substitutions, irradiation or strong magnetic field [16] has been attributed to a metal insulator transition or scattering on magnetic defects but the detailed microscopic mechanisms remain unclear.

To reveal the low-T resistivity upturn $\Delta\rho(T) = \rho(T) - \rho_m(T)$, we separated it from the resistivity $\rho_m(T) = \rho_0 + \sum_{s=1}^{s=m} a_s T^s$ at high T ~ 100-400K. Irrespective to the particular form of $\rho_m(T)$ (usually the quadratic polynomial works well), $\Delta\rho(T)$ above $T_c$ displays the logarithmic behaviour, the magnitude of $\Delta\rho(T)$ increasing with the dose. This is shown in Fig.2 where $\Delta\rho(T) = \rho(T) - \rho_2(T)$ is plotted for each irradiation step and fitted by the equation:

$$\Delta\rho(T) = A_K \ln\left[1 + (T/T_K)^n\right] \qquad (1)$$

Here 1 in the brackets was added so that $\Delta\rho(T)$ decreases rapidly at $T > T_K$ and $\Delta\rho(T)$ can be separated from $\rho_0$. All $\Delta\rho(T)$ curves can be well fitted by Eq.(1) from about 5-10 K above $T_c$ to 60-80 K taking $A_K$ and $T_K$ as fit parameters and adjusting the exponent $n$ to describe the data above 90 K. For the highest dose, $\Delta\rho(T)$ remains logarithmic down to 2K. The inset of Fig. 2 shows that $A_K$ increases monotonically with the fluence and nearly triples from the first to the last irradiation. In turn, $T_K$ does not vary significantly remaining between 106 K and 119 K, while the exponent $n$ varies from 5 to 3. As shown in Fig. 3, both the residual resistivity $\rho_0$ and $\rho_{50K}$ at 50 K extracted from the high-T polynomial fit increase in the same way with the fluence while the ratios $\Delta\rho(T)/\rho_0$ for all irradiation doses collapse onto a single universal curve.

The totality of our data, namely the logarithmic temperature dependence of $\Delta\rho(T)$, the increase of $A_k$ with the fluence, the independence of $T_K$ of the fluence suggests that irradiation defects have uncompensated spins, which cause Kondo-like scattering. The significant increase of the residual resistivity $\rho_0$ indicates that irradiation produces nonmagnetic scatters as well. Moreover, the fact that the ratio $\Delta\rho(T)/\rho_0$ for all irradiation doses collapses onto a single universal curve implies that both $\Delta\rho(T)$ and $\rho_0$ have the same dependence on the concentration of the irradiation defects. Thus, the irradiation defects cause both magnetic and nonmagnetic scattering.

Logarithmic dependence of $\Delta\rho(T)$ was observed on electron irradiated underdoped YBCO [14] for which $T_K$ is similar to Nd-1111. However, $A_K$ for Nd-1111 is more than twice of that for YBCO, consistent with the idea that displacements of Nd and Fe ions by $\alpha$ particles can produce uncompensated magnetic moments. For the Kondo scenario, the $lnT$ dependence of $\Delta\rho(T)$ tends to saturate at low T and it is suppressed by magnetic fields. In cuprates the saturation of $\Delta\rho(T)$ has been observed for $T/T_K \sim 0.01$ [17] and in strong fields up to 50 T [18]. At our lowest temperature of 2 K ($T/T_K \sim 0.02$) we only observed a slight flattening of $\Delta\rho(T)$, and a negative magnetoresistivity $[\rho(9T)-\rho(0T)]/\rho(0T) \approx -0.05$ consistent with the low-T upturn $\Delta\rho(T)$ caused by the spin-flip scattering.

As shown in Figs. 2 and 3, neither $A_K$ nor $\rho_0$ nor $\rho_{50K}$ increase linearly with the fluence. This may be due to the well-established effect of partial annihilation of defects generated during previous doses or a shift of the chemical potential due to irradiation discussed below. The quantity $\Delta\rho_0 = \rho_0^{(i)} - \rho_0^{unirr}$ where $\rho_0^{(i)}$ is the residual resistivity after i-th irradiation can be used as a measure of concentration of irradiation defects. Shown in Fig. 4 is the dependence $T_c(\Delta\rho_0)$ which exhibits a linear decrease at small $\Delta\rho_0$ followed by a sharper drop at higher $\Delta\rho_0$. Since irradiation can also shift the chemical potential, we measured the Hall resistivity $R_H(T)$ which shows that the effective carrier density $n=1/eR_H$ increases linearly with irradiation (less than by a factor 2 for the maximum dose). In multiband pnictids $R_H(T)$ may not be a good measure of the carrier density, yet re-plotting $T_c$ as a function of $\Delta\rho_H = \Delta\rho_0 R_H/R_H^{(i)}$ in which only the mean free path is affected by irradiation, makes $T_c(\Delta\rho_H)$ nearly linear and doubles the critical value of $\Delta\rho_H$ at which $T_c(\Delta\rho_H)$ goes to zero.

The nonmagnetic scattering rate $\Gamma$ by irradiation defects can be evaluated qualitatively from the relation $\Gamma = \Delta\rho_0/\mu_0\lambda_0^2$ assuming that intraband contributions from the electron and hole bands and interband scattering rates are of the same order of magnitude. Here $\lambda_0 = 195$ nm is the London penetration depth for an unirradiated Nd-1111 crystal at $T = 0$ [19]. The dimensionless interband scattering rate $g = \Gamma\hbar/4\pi k_B T_{c0} \cong \hbar\Delta\rho_0/4\pi k_B T_{c0}\mu_0\lambda_0^2$ which defines the pairbreaking effect in multiband models [7] is shown on the upper axes in Fig. 4. The $g$ parameter varies from 0 to 1.7 and its maximum value nearly doubles if the increase of the carrier density with disorder is taken into account.

As our data indicate, irradiation defects produce strong magnetic and nonmagnetic disorder so the contribution to $g$ from the spin flip and nonmagnetic scattering may be of the same order of magnitude. These g values are much larger than the critical $g_c$ at which $s^{\pm}$ superconductivity is destroyed by impurities treated in the Born approximation. For equal gaps $\Delta$ of opposite signs on different sheets of the Fermi surface, this theory gives the equation $ln(T_{c0}/T_c) = \psi(1/2+g) - \psi(1/2)$ in which $T_c$ vanishes above the critical value $g_c \approx 0.15$. Here $\psi(x)$ is the di-gamma function, $T_{c0}$ is the critical temperature before irradiation, $g = (\Gamma_s^{intra} + \Gamma_n^{inter})\hbar/4\pi k_B T_{c0}$ [7] where $\Gamma_s^{intra}$ and $\Gamma_n^{inter}$ are the spin flip intraband and the nonmagnetic interband scattering rates, respectively. This estimate of $g_c$ is more than 10 times smaller than the experimental value of $g_c \approx 1.5$ (or 20 times smaller if the change of the carrier density due to irradiation is taken into account). The large values of g result from a high density of irradiation defects: the estimate of the mean free path $l = v_F/\Gamma$ with the Fermi velocity $v_F = 1.3 \times 10^5$ m/s [20] gives $l \approx 2.4$ nm (or ~1 nm given the increase of the carrier density). It is rather remarkable that multiband superconductivity in Nd-1111 turns out to be so resilient to such strong and dense magnetic and nonmagnetic disorder. In that respect Nd-1111 behaves more like conventional s-wave $MgB_2$ and $V_3Si$ with no sign change of the order parameter rather than the d-wave cuprates like $YBa_2Cu_3O_7$, as shown in the inset of Fig. 4.

The results presented above show inconsistencies of our data with multiband BCS models, which assume uncorrelated weak scattering by dilute impurities. It has been shown recently that strong single impurity scattering can reduce pairbreaking by static disorder in multiband models [7]. Yet the fact that the density of irradiation defects in our experiments is so high that they start changing the Hall resistivity suggests that spatial correlations of impurity scattering may become important. It is also puzzling that, for such apparently very dense structure of magnetic scatterers, $\Delta\rho(T) \propto lnT$ shows no apparent signs of saturation due to interaction of magnetic moments [18,21]. Other possibilities for the logarithmic $\Delta\rho(T)$, such as the 2D weak localization or granularity effects [22] are less feasible given the moderate mass anisotropy [11] and the lack of grain boundaries in our Nd-1111 single crystals.

In conclusion, we report the effect of α-particle irradiation on superconducting and nonsuperconducting properties of Nd-1111 single crystals. Our results indicate that irradiation defects produce both spin-flip and nonmagnetic scattering yet multiband superconductivity in pnictides survives up to an unusually high concentration of irradiation defects for which magnetic intraband scattering and nonmagnetic interband scattering would be expected to suppress $T_c$.

The work was supported by the NSF grant DMR-0084173, the state of Florida and AFOSR (FA9550-06-1-0474), Italian Foreign Affairs Ministry, General Direction for the Cultural Promotion. AG is grateful to O. Dolgov, P.J. Hirschfeld and D.J. Scalapino for discussions and to KITP at UCSB where a part of this work was completed under support of NSF Grant No. PHY05-51164.

**Figure Captions**

Fig. 1. $\rho(T)$ measured after each step of irradiation. Inset shows magnified temperature region near $T_c$.

Fig. 2. $\Delta\rho(T)$ curves after each irradiation step. The best fits to Eq. (1) for the highest and lowest doses are shown by continuous black lines. Inset shows $A_K$ as a function of the fluence.

Fig. 3. The ratio $\Delta\rho(T)/\rho_0$ for all irradiation doses. Inset shows $\rho_0$ (left y-axis) and $\rho_{50K}$ (right y-axis) as functions of the fluence.

Fig. 4. $T_c$ versus $\Delta\rho_0$ and $g = \Gamma\hbar/4\pi k_B T_{c0}$. Inset shows comparison of $T_c$ suppression by irradiation for Nd-1111 (this work), YBCO [17], MgB$_2$ and V$_3$Si [23].

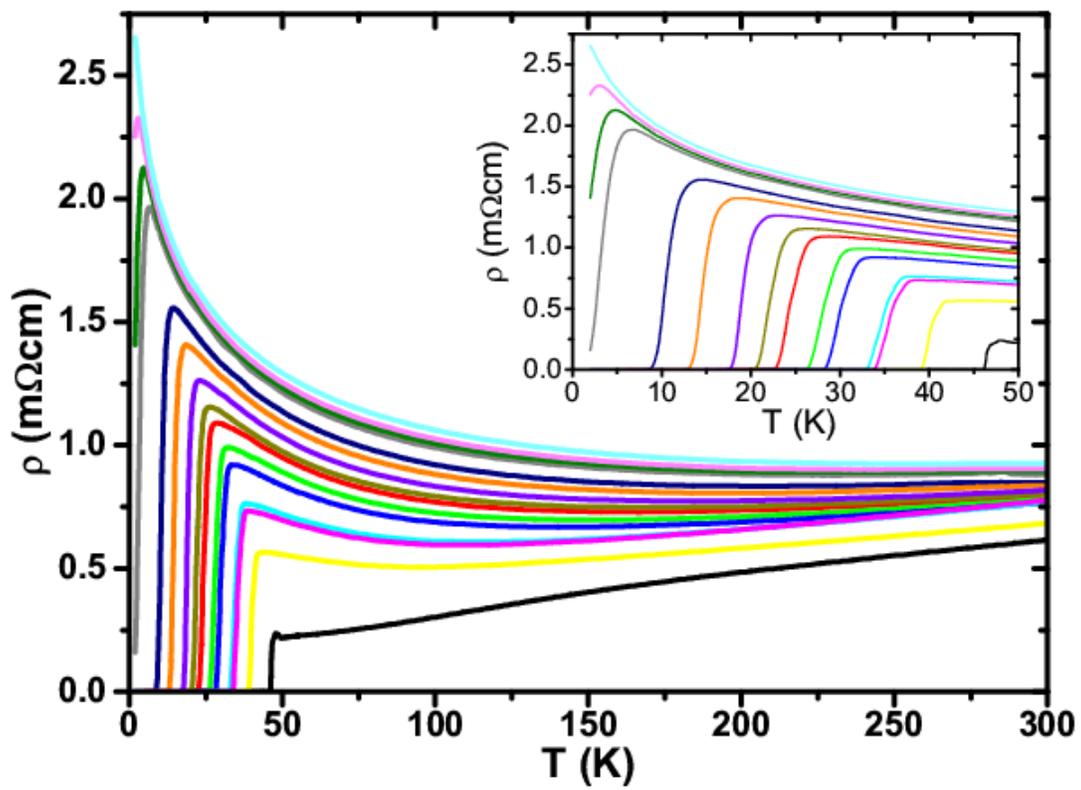

**Figure 1**

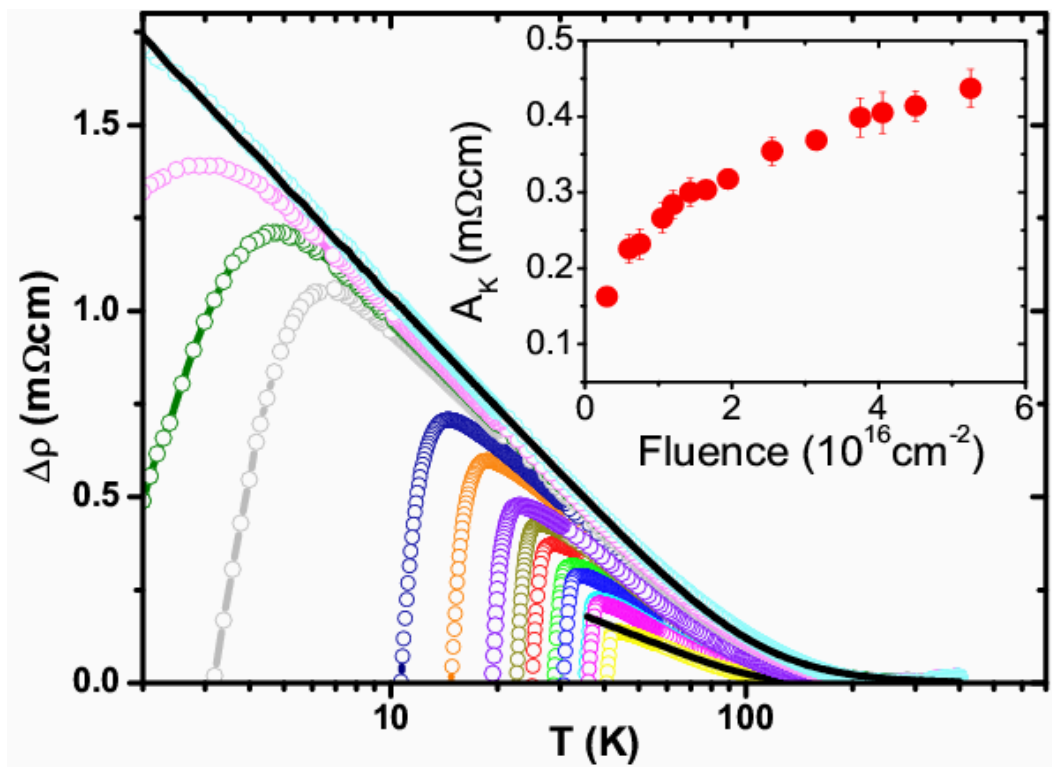

**Figure 2**

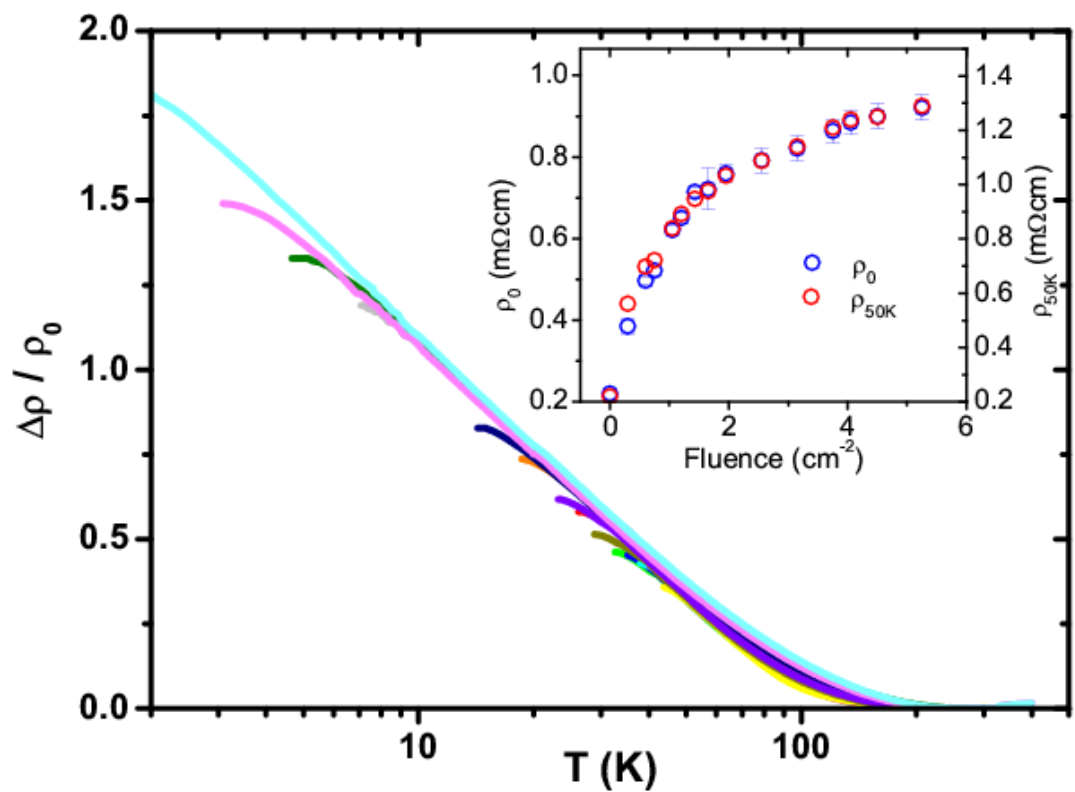

**Figure 3**

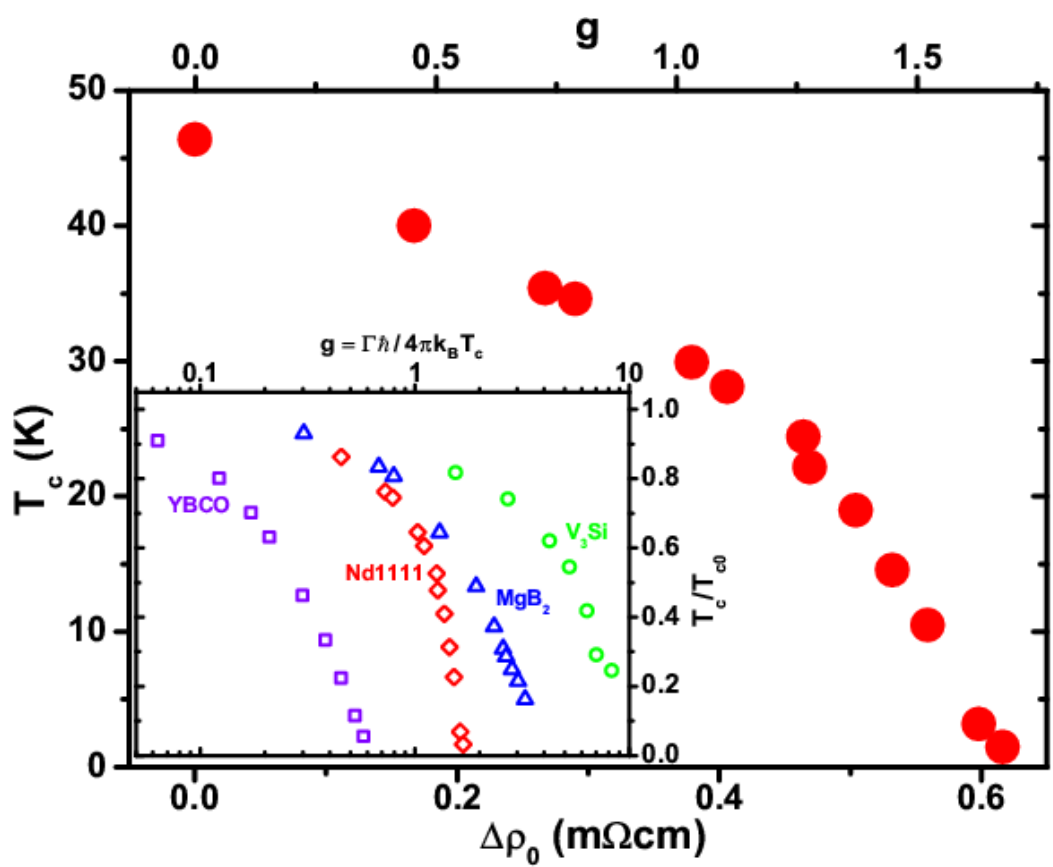

**Figure 4**